\documentclass{kapproc} 

\usepackage{procps} 

\usepackage[dvips]{graphicx}

\upperandlowercase

 \nochapequationnumber 

\normallatexbib 

\begin{document}
\articletitle{Chaos and Quantum Mechanics}

\author{Salman Habib}
\affil{MS B285, Theoretical Division, The University of California, 
Los Alamos National Laboratory, Los Alamos, New Mexico
87545} 
\email{habib@lanl.gov}

\author{Tanmoy Bhattacharya,\altaffilmark{1} Benjamin
Greenbaum,\altaffilmark{2} Kurt Jacobs,\altaffilmark{1,3} \\ Kosuke
Shizume,\altaffilmark{4} and Bala Sundaram,\altaffilmark{5}}
   
\affil{\altaffilmark{1}Los Alamos National Laboratory,
\altaffilmark{2}Columbia University, \altaffilmark{3}Griffith
University, \altaffilmark{4}Tsukuba University, \altaffilmark{5}City
University of New York}  

\begin{abstract}

The relationship between chaos and quantum mechanics has been somewhat
uneasy -- even stormy, in the minds of some people. However, much of
the confusion may stem from inappropriate comparisons using formal
analyses. In contrast, our starting point here is that a complete
dynamical description requires a full understanding of the evolution
of {\em measured systems}, necessary to explain actual experimental
results. This is of course true, both classically and quantum
mechanically. Because the evolution of the physical state is now
conditioned on measurement results, the dynamics of such systems is
intrinsically nonlinear even at the level of distribution
functions. Due to this feature, the physically more complete treatment
reveals the existence of dynamical regimes -- such as chaos -- that
have no direct counterpart in the linear (unobserved) case. Moreover,
this treatment allows for understanding how an effective classical
behavior can result from the dynamics of an observed quantum system,
both at the level of trajectories as well as distribution
functions. Finally, we have the striking prediction that time-series
from measured quantum systems can be chaotic far from the classical
regime, with Lyapunov exponents differing from their classical
values. These predictions can be tested in next-generation
experiments.

\end{abstract}

\section{Prologue}

I met Henry Kandrup as a graduate student at Maryland in 1985, having
recently decided to switch from experiment to theory. My first
interaction with postdocs -- at the time an intimidatingly higher
form of life -- occurred when Henry suggested that he and another
postdoc, Ping Yip, and I take up the question of Landau damping and
stability of star clusters. While I was happy to work with Henry and
Ping, most of the time I was struggling to understand cryptic
conversations laced with mathematical jargon -- ``functions of compact
support,'' ``consider the following inner product,'' and so on. Since
I wasn't following too much of this, I decided it was better to go
away and catch up by reading every paper that was even vaguely related
to the topic. This turned out to be much easier than expected, and one
night I came up with a simple way of combining Henry's previous work
on stability with conventional Landau damping theory from plasma
physics. Coming in to the department late in the morning I showed the
first set of notes to Henry. He looked at them, did not say much --
which was unusual -- and went back home. Next day, as I entered my
office, I was stunned to find, slipped under the door, a complete
preprint of a paper, all equations written in by hand in Henry's
beautiful copperplate. He had gone home, generalized my notes to the
problem at hand, worked through the entire thing, come back late at
night, and typed the preprint on an electric typewriter (this was just
before the advent of word processing), finishing as the sun came
up. After this incident, I was {\em really} afraid of postdocs!

My early interactions with Henry were very wide-ranging; we discussed
all sorts of topics, from classical statistical mechanics to quantum
gravity, and on all of them he was very well-informed and
entertainingly opinionated. The years went by quickly, Henry moved on
to other places and so did I. Although we argued and collaborated now
and then as of old, in my memory the early years have a certain
luminescence. My favorite remembrance of Henry is that after he had
demolished somebody's hapless piece of research in one of our
discussions, he would look up, smile in a disarming way, and say,
``True?'' It usually was.

One of the topics Henry and I discussed at considerable length and
depth was the nature of chaos in multi-particle systems and its role
in controlling aspects of the dynamical behavior of statistical
averages. While we did not always agree, these discussions certainly
attuned my thinking about the problem. In this contribution, I present
a discussion of how to think about chaos in a physical way, from the
point of view of realistic experiments. The basis of the arguments
applies to both classical and quantum systems and serves to bring
together these two great dynamical traditions that are seemingly at
such odds with each other. The work reported here is the result of
several collaborations between subsets of the authors.  While I do not
know what Henry's opinions would have been on this subject, however, I
am sure he would not have been quiet!

\section{Introduction}

In classical theory -- unlike in quantum mechanics -- the status of
dynamical chaos is apparently clear: chaos exists observationally and
is well-described theoretically by Newton's equations. (Nevertheless,
even here, a deeper look at the physical meaning of chaos is certainly
helpful; we return to this presently.) It is in the context of
quantum theory, however, that the notion of chaos appears so puzzling
and mysterious. Because of the Kosloff-Rice theorem~\cite{kr} and
related results~\cite{recur}, it is clear that quantum evolution of
the wave function or the density matrix is integrable; hence, chaos
cannot exist in quantum mechanics in the canonical sense. This is the
basic stumbling block to defining a quantum notion of
nonintegrability.

One may argue that real quantum systems are always coupled to an
environment and hence their evolution -- ``for all practical
purposes,'' (FAPP), in Bell's famous phrase~\cite{bell} -- should be
described by unitarity-breaking master equations rather than the
unitary evolution assumed by the Kosloff-Rice theorem. Perhaps this
way out, although not fundamentally satisfying to the purist, is enough
by itself, but it is easy to see what is wrong with the
argument. Fundamentally, any fully quantum dynamical description must
arise from a Hamiltonian describing the system, its environment, and
their coupling. The master equation represents the evolution of the
reduced density matrix for the system which arises from tracing over
the environment variables in the full (system plus environment)
density matrix. Since the full evolution must satisfy Kosloff-Rice,
the evolution of the reduced density matrix cannot be nonintegrable.

Thus, the fundamental problem we are faced with is this: we are
familiar with chaos in the real world, but our fundamental theory of
dynamics -- which passes every experimental test beautifully --
seemingly does not have a natural place within it to tolerate even the
existence of the concept. This should not come as a surprise;
after all, the trajectories of classical mechanics are apparently
``real'' and effortless to contemplate, but they too, have no natural
place in quantum mechanics. Now it is true that quantum mechanics is
an intrinsically probabilistic theory, but that, in itself, is not the
real issue. Classical theory can be easily cast as fundamentally
probabilistic as well, via the classical Liouville equation describing
the evolution of a classical probability in phase space. (For an
attempt at an even closer analogy, see Ref.~\cite{koop} and the
discussion in Ref.~\cite{peres}.) As discussed further below, the key
point is rather that, unlike special relativity, where $v/c\rightarrow
0$ smoothly transitions between Einstein and Newton, the limit
$\hbar\rightarrow 0$ is singular. The symmetries underlying quantum
and classical dynamics -- unitarity and symplecticity, respectively --
are fundamentally incompatible with the opposing theory's notion of a
physical state: quantum-mechanically, a positive semidefinite density
matrix; classically, a positive phase-space distribution function.

In the rest of this article, we will expose the singular nature of the
$\hbar\rightarrow 0$ limit and discuss a physical point of view --
applicable to both classical and quantum systems -- which will enable
us to explain how trajectories and chaos appear in real experiments.

At this point, it should be clear that the questions taken up in this
contribution are not those usually considered under the research area
called ``quantum chaos.'' There, one is primarily interested in the
quantum behavior of a system with a classically chaotic Hamiltonian,
what might happen to the validity of certain approximations (e.g.,
semiclassical approaches to calculating the quantum propagator) and
whether classical trajectories and phase space structures can provide
some insight into the nature of quantum wavefunctions. But one does
not actually study quantum chaos.

We distinguish between {\em isolated} evolution, where the system
state evolves without any coupling to the external world, {\em
unconditioned open} evolution, where the system evolves coupled to an
external environment but where no information regarding the system is
extracted from the environment, and {\em conditioned open} evolution
where such information {\em is} extracted. In the third case, the
evolution of the physical state is driven by the system evolution, the
coupling to the external world, and by the fact that observational
information regarding the state has been obtained. This last aspect --
system evolution {\em conditioned} on the measurement results via
Bayesian inference -- leads to an intrinsically nonlinear evolution
for the system state, and distinguishes it from unconditioned
evolution. While the concept of conditioned evolution of the system
state is familiar to engineers and mathematicians, especially systems
engineers and control theorists~\cite{control}, it is not yet
completely familiar territory to the majority of
physicists. Nevertheless, driven by the impressive progress in the
experimental state-of-the-art in quantum and atomic optics and in
nanoscience~\cite{exp}, these notions are now being employed as
everyday tools at least in some fields.

The conditioned evolution provides, in principle, the most realistic
possible description of an experiment. To the extent that quantum and
classical mechanics are eventually just methodological tools to
explain and predict the results of experiments, this is the proper
context in which to compare them and discuss the nature of predictions
for real experiments. The explicit incorporation of information gained
via measurement also provides a structure to address the
quantum-classical transition more generally, and to frame the question
of where chaos exists within this structure.

The fact that quantum and classical mechanics are fundamentally
incompatible in many ways, yet the macroscopic world is well-described
by classical dynamics has puzzled physicists ever since the laying of
the foundations of quantum theory. It is fair to say that not everyone
is satisfied with the state of affairs -- including many seasoned
practitioners of quantum mechanics.

Of course, the notion of measurement in quantum mechanics -- the
denial of reality to system properties unless they are measured -- is
such a revolutionary concept that it engenders much more
unease~\cite{bell2}, even today. The problem is that, were quantum
mechanics the final theory, it could deny reality to the measurement
results themselves unless they were observed by another system and so
on, {\em ad infinitum}. In order to ``solve'' the ``measurement
problem,'' it originally appeared impossible to think of quantum
mechanics as a fundamental theory without relying on the existence of
a classical world-view within which to embed it~\cite{llqm}. Although
we still cannot dispel the unease invoked by the measurement problem,
it is important to stress that the quantum-classical transition can be
understood independently. This transition should not be confused with
the measurement problem.

A partial understanding of the classical limit arises from the idea --
familiar from nonequilibrium statistical mechanics -- that weak
interactions of a system with an environment are
universal~\cite{llsm}. These interactions can effectively suppress
certain nonclassical terms in the quantum
evolution~\cite{deco}. However, at best they only allow for the
emergence of a classical probabilistic evolution and it can be shown
that the mere existence of such interactions is insufficient to yield
classical evolution in all cases~\cite{shetal}. Finally, this picture
alone cannot explain the results of actual measurements where
information can be continuously extracted from the environment and
used to define operational notions of a trajectory. We now go in to
these questions in more detail.

\section{Isolated and Open Evolution}

Suppose we are given an arbitrary system Hamiltonian $H(x,p)$ in terms
of the dynamical variables $x$ and $p$; we will be more specific
regarding the precise meaning of $x$ and $p$ as position and momentum
later. The Hamiltonian is the generator of time evolution for the
physical system state, provided there is no coupling to an environment
or measurement device. In the classical case, we specify the initial
state by a positive phase space distribution function $f_{Cl}(x,p)$;
in the quantum case, by the (position-representation) positive
semidefinite density matrix $\rho(x_1,x_2)$ or, completely
equivalently, by the corresponding Wigner distribution function
$f_W(x,p)$ (not positive). The Wigner distribution~\cite{wdf,sh90} is a
``half-Fourier'' transform of $\rho(x_1,x_2)$, defined as
\begin{equation}
f_W(x,p)=\frac{1}{2\pi\hbar}\int d\Delta 
\rho(x+\frac{1}{2}\Delta,x-\frac{1}{2}\Delta)\exp(-ip\Delta/\hbar),
\label{wdef}
\end{equation}
where $x\equiv (x_1+x_2)/2$ and $\Delta\equiv x_1-x_2$.

The evolution of an {\em isolated} system is then given by the
classical and quantum Liouville equations for the {\em fine-grained}
distribution functions (i.e., the evolution is entropy-preserving):
\begin{equation} 
\partial_t f_{Cl}(x,p)=-\left[\frac{p}{m}\partial_x - 
\partial_x V(x) \partial_p \right] f_{Cl}(x,p),
\label{cle}
\end{equation}
\begin{eqnarray} 
\partial_t f_W(x,p)&=&-\left[\frac{p}{m}\partial_x - 
\partial_x V(x) \partial_p \right] f_W(x,p) \nonumber\\ 
&& + \sum_{\lambda = 1}^\infty\frac{(\hbar/2i)^{2\lambda}}{(2\lambda + 1)!} 
\partial_x^{2\lambda+1}V(x) \partial_p^{2\lambda+1} f_W(x,p),
\label{qle} 
\end{eqnarray}
where we have assumed for simplicity that the potential $V(x)$ can be
Taylor-expanded; this does not alter the nature of any of the
following arguments. Note that these evolutions are both linear in the
respective distribution functions. 

The limiting form $f_{Cl}(x,p)=\delta(x-\bar{x})\delta(p-\bar{p})$ is
allowed classically, and, on substitution in Eqn.~(\ref{cle}), yields
the expected Newton's equations. These may then be interpreted as
equations for the particle position and momentum, although we must
emphasize that this identification is only formal at this stage.
Quantum mechanically, this ultralocal limit is not permitted since
$f_W(x,p)$ must be square-integrable, therefore -- even formally -- no
direct particle interpretation can exist. In both cases, if one allows
for initially localized distributions but which nevertheless have some
finite width, it is easy to see that if $V(x)$ is nonlinear, quite
generically the distribution will eventually spread over the allowed
phase space and not remain localized.

As alluded to in the Introduction, the extension to open systems is
conceptually trivial, but very difficult to implement in practice. To
the original system Hamiltonian, we now add pieces representing the
environment and the system-environment coupling. If the environment is
in principle unobservable, then a (nonlocal in time) linear master
equation for the system's reduced density matrix is -- in theory --
derivable by tracing over the environmental variables. In practice,
tractable equations are impossible to obtain without drastic
simplifying assumptions such as weak coupling, timescale separations,
and simple forms for the environmental and coupling Hamiltonians. In
any case, the important point to note is that the act of tracing over
the environment does not change the linear nature of the
equations. Generally speaking, master equations describing open
evolution of {\em coarse-grained} distributions augment the RHS of
Eqns.~(\ref{cle}) and (\ref{qle}) with terms containing dissipation
and diffusion kernels connected via generalized
fluctuation-dissipation relations~\cite{noneq}. While the classical
diffusion term vanishes in the limit of zero temperature for the
environment, this is not true quantum mechanically due to the presence
of zero-point fluctuations.

\section{Continuous Measurement and Conditioned Evolution}

In contrast to classical theory, where measurement can be, in
principle, a passive process, in quantum theory measurement creates an
irreducible disturbance on the observed system (quantum
``backaction''). This being so, if our aim is that measurement yield
dynamical information -- rather than strongly influence dynamics --
the desired measurement process must yield a limited amount of
information in a finite time. Hence, simple projective (von Neumann)
measurements are clearly not appropriate because they yield complete
information instantaneously via state projection. Nevertheless, this
fundamental notion of measurement can be easily extended~\cite{peres}
to devise schemes that extract information
continuously~\cite{cmeqns}. The basic idea is to have the system of
interest interact weakly with another (e.g., atom interacting with an
electromagnetic field) and make projective measurements on the
auxiliary system (e.g., photon counting). Because of the weak
interaction, the state of the auxiliary system gathers very little
information regarding the system of interest, and therefore this
system, in turn, is only perturbed slightly by the measurement
backaction. Only a small component of the information gathered by the
projective measurement of the auxiliary system relates to the system
of interest, and a continuous limit of the measurement process can be
taken.

In the continuous limit, the evolution of the system density matrix is
fundamentally different from the equations discussed above for the
case of open evolution. The master equation describing the evolution
of the reduced density matrix conditioned on the results of the
measurements contains a term that reflects the gain in information
arising from the measurement record (``innovation'' in the language of
control theory). This term, arising from applying a continuous analog
of Bayes' theorem, is intrinsically nonlinear in the distribution
function. The coupling to an external probe (and the associated
environment) will also cause effects very similar to the open
evolution considered earlier, and there can once again be dissipation
and diffusion terms in the evolution equations. The primary
differences between the classical and quantum treatments, aside from
the kinematic constraints on the distribution functions, are the
following: (i) the (nonlocal in $p$) quantum evolution term in
Eqn.~(\ref{qle}), and (ii) an irreducible diffusion contribution due
to quantum backaction reflecting the {\em active} nature of quantum
measurements.

We now consider a simple model of position measurement to provide a
measure of concreteness. In this model, we will assume that there are
no environmental channels aside from those associated with the
measurement. Suppose we have a single quantum degree of freedom,
position in this case, undergoing a weak, ideal continuous
measurement~\cite{cmeqns}. Here ``ideal'' refers to no loss of
information during the measurement, i.e., a fine-grained evolution
with no increase in entropy. Then, we have two coupled equations, one
for the measurement record $y(t)$,
\begin{equation}
dy = \langle x \rangle dt + \frac{1}{\sqrt{8k}}dW
\label{record}
\end{equation}
where $dy$ is the infinitesimal change in the output of the
measurement device in time $dt$, the parameter $k$ characterizes the
rate at which the measurement extracts information about the
observable, i.e., the {\em strength} of the measurement~\cite{djj},
and $dW$ is the Wiener increment describing driving by Gaussian white
noise~\cite{noise}, the difference between the actually observed value
and that expected. The other equation -- the nonlinear stochastic master
equation (SME) -- specifies the resulting conditioned evolution of the
system density matrix, given in the Wigner representation,
\begin{eqnarray}
f_W(x,p,t+dt)&=&\left[ 1 + dt\left[-\frac{p}{m} \partial_x  + \partial_x
V(x)\partial_p + D_{BA}\partial_p^2\right]\right. \nonumber\\
&& + \left.dt\sum_{\lambda = 1}^\infty \! \frac{\left(
\hbar/2i\right)^{2\lambda}}{(2\lambda + 1)!}  \partial_x^{2\lambda+1} 
V(x,t) \partial_p^{2\lambda+1} \right] f_W(x,p,t)
\nonumber \\
&& + dt\sqrt{8k}(x - \langle x \rangle)f_W(x,p,t)dW, 
\label{condq}
\end{eqnarray} 
where $D_{BA}=\hbar^2k$ is the diffusion coefficient arising from
quantum backaction and the last (nonlinear) term represents the
conditioning due to the measurement. In principle, there is also a
(generalized) damping term~\cite{mm}, but if the measurement coupling
is weak enough, it can be neglected. If we choose to average over all
the measurement results, which is the same as ignoring them, then the
conditioning term vanishes, but {\em not} the diffusion from the
measurement backaction. Thus the resulting linear evolution of the
coarse-grained quantum distribution is not the same as the linear
fine-grained evolution~(\ref{qle}), but yields a conventional
open-system master equation. Moreover, for a given (coarse-grained)
master equation, different underlying fine-grained SME's may exist,
specifying different measurement possibilities.

In contrast to the quantum case, the corresponding (ideal) classical
conditioned master equation [set $\hbar=0$ in Eqn.~(\ref{condq}),
holding $k$ fixed],  
\begin{eqnarray}
f_{Cl}(x,p,t+dt)&=&\left[ 1 - dt\left[\frac{p}{m} \partial_x  -
\partial_x V(x)\partial_p \right]\right]f_{Cl}(x,p,t) \nonumber\\
&& + dt \sqrt{8k}(x - \langle x \rangle)f_{Cl}(x,p,t)dW,
\label{condc}
\end{eqnarray}
does not have the backaction term as these classical measurements are
{\em passive}: averaging over all measurements simply gives back the
Liouville equation (\ref{cle}), and there is no difference between the
fine-grained and coarse-grained evolutions in this special case. [In
general, classical diffusion terms from ordinary open evolution can
also coexist, as in the {\em a posteriori} evolution specified by the
Kushner-Stratonovich equation~\cite{kse}, of which Eqn.~(\ref{condc})
is a special case.] As a final point, we delay our discussion of how
the classical trajectory limit is incorporated in Eqn.~(\ref{condc}),
i.e., the precise sense in which the ``the position of a particle is
what a position-detector detects,'' to the next section.

\section{QCT: The Quantum-Classical Transition}

If quantum mechanics is really the fundamental theory of our world,
then an effectively classical description of macroscopic systems must
emerge from it -- the so-called quantum-classical transition (QCT). It
turns out that this issue is inextricably connected with the question
of the physical meaning of dynamical nonlinearity discussed
above. Having written down the relevant evolution equations, we now
analyze two notions of the QCT and how they emerge from the equations.

Quantum mechanics is intrinsically probabilistic, but classical theory
-- as shown above by the existence of the delta-function limit for the
classical distribution function -- is not. Since Newton's equations
provide an excellent description of observed classical systems,
including chaotic systems, it is crucial to establish how such a
localized, or trajectory, description can arise quantum
mechanically. We will call this the {\em strong} form of the QCT. Of
course, in many situations, only a statistical description is possible
even classically, and here we demand only the agreement of quantum and
classical distributions and the associated dynamical averages. This
defines the {\em weak} form of the QCT.

It is clear that if the strong form of the QCT holds, then, via
trivial coarse-graining, the weak form follows automatically. The
reverse is not true, however: results from a coarse-grained analysis
cannot be applied to the fine-grained situation. Moreover, the
violation of the conditions necessary to establish the strong form of
the QCT need not prevent the existence of a weak QCT. We now discuss
and establish the conditions under which these transitions
occur. Since the strong form of the QCT requires treating the
localized limit, a cumulant expansion for the distribution function
immediately suggests itself, whereas, for the more nonlocal issues
relevant to the weak form of the QCT, a semiclassical analysis turns
out to be natural.

\subsection{Strong Form of the QCT: Chaos in the Classical Limit}

It is easy to see that the strong form of the QCT is impossible to
obtain from either the isolated or open evolution equations for the
density matrix or Wigner function. As mentioned already, for a generic
dynamical system, a localized initial distribution tends to distribute
itself over phase space -- and then continue to evolve -- either in
complicated ways (isolated system) or asymptote to an equilibrium
state (open system), whether classically or quantum mechanically. In
the case of conditioned evolution, however, the distribution can be
localized due to the information gained from the measurement, and
evolve in a quite different manner. In order to quantify how this
happens, let us first apply a cumulant expansion to the (fine-grained)
conditioned classical evolution~(\ref{condc}). This results in the
following equations for the centroids ($\bar{x}\equiv\langle
x\rangle$, $\bar{p}\equiv\langle p\rangle$),
\begin{eqnarray}
d\bar{x}&=&\frac{\bar{p}}{m}dt+\sqrt{8k}C_{xx}dW,\nonumber\\
d\bar{p}&=&\langle F(x)\rangle dt+\sqrt{8k}C_{xp}dW,
\label{cumc}
\end{eqnarray}
where
\begin{eqnarray}
F(x)&=&-\partial_x V(x),\nonumber\\
C_{AB}&=&\frac{1}{2}(\langle AB\rangle + \langle
BA\rangle -2\langle A\rangle\langle B\rangle),
\end{eqnarray}
along with a hierarchy of coupled equations for the time-evolution of
the higher cumulants. These equations are the continuous measurement,
real-world, analog of the formal ultralocal Newtonian limit of the
distribution function in the classical Liouville
equation~(\ref{cle}). Whereas Eqns.~(\ref{cumc}) always apply, our aim
is to determine the conditions under which the cumulant expansion
effectively truncates and brings their solution very close to that of
Newton's equations.  This will be true provided the noise terms are
small (in an average sense) and the force term is localized, i.e.,
$\langle F(x)\rangle=F(\bar{x})+\cdots$, the corrections being
small. The required analysis involves higher cumulants and has been
carried out elsewhere~\cite{bhj}. It turns out that the distribution
is localized provided
\begin{equation}
8k\gg\sqrt{\frac{(\partial_x^2 F)^2|\partial_x F|}{2mF^2}}
\label{klarge}
\end{equation}
and the motion of the centroid will effectively define a smooth
classical trajectory -- the low-noise condition -- as long as
\begin{equation}
k\gg \frac{2|\partial_x F|}{S}
\end{equation}
where $S$ is the action scale of the system. Note that this condition
does not bound the measurement strength: classically we can always
extract as much information as needed -- at least in principle -- to
gain the trajectory limit. This, then, is the ``realistic'' derivation
of Newton's equations.

We now turn to the quantum version of these results. In this case, the
analogous cumulant expansion gives exactly the same equations for the
centroids as above, while the equations for the higher cumulants are
different. (The evolution of classical and quantum averages is the
same to Gaussian order, with the first differences arising at the next
order~\cite{shgauss}.) We can again investigate whether a trajectory
limit exists. Localization holds in the weakly nonlinear case if the
classical condition above is satisfied. In the case of strong
nonlinearity, the inequality becomes~\cite{bhj}
\begin{equation}
8k\gg\frac{(\partial_x^2 F)^2\hbar}{4mF^2}.
\end{equation}
Because of the backaction, the low-noise condition is implemented in
the quantum case by a double-sided inequality:
\begin{equation}
\frac{2|\partial_x F|}{s}\ll\hbar k\ll\frac{|\partial_x F|s}{4},
\label{newtonq}
\end{equation}
where the action is measured in units of $\hbar$, $s$ being
dimensionless. The left inequality is the same as the classical one
discussed above, however the right inequality is essentially quantum
mechanical. The measurement strength cannot be made arbitrarily large
as the backaction will result in too large a noise in the equations
for the centroids. As the action $s$ is made larger, both inequalities
are satisfied for an ever wider range of values of $k$. For
sufficiently large $s$, the actual value of $k$ becomes irrelevant and
the dynamics becomes effectively classical.

To recapitulate, for continuously measured quantum systems,
trajectories that emerge in the macroscopic limit follow Newton's
equations, and hence can be chaotic as shown
elsewhere~\cite{bhj}. Thus, as speculated in a prescient paper by
Chirikov~\cite{chirikov}, measurement indeed provides the missing link
between ``quantum'' and ``chaos,'' at least in the classical limit.

Finally, in experiments one usually considers the measurement record
itself rather than the estimated state of the system as we have
discussed so far.  As measurement introduces a white noise, it is
important to investigate the condition under which the record tracks
the estimate faithfully.  If $\Delta t$ is the time over which the
continuous measurement is averaged to obtain the record (this
averaging being a necessary part of any finite-bandwidth experiment),
and we allow ourselves a maximum of $\Delta x$ as the position noise,
it is easy to see that the measurement strength needs to
satisfy~\cite{bhj} 
\begin{equation}
8k > \frac{1}{\Delta t (\Delta x)^2}
\label{kobs}
\end{equation}

To demonstrate these results for a concrete example, we revisit the
results of Ref.~\cite{bhj} for a driven, Duffing oscillator, with
system Hamiltonian
\begin{equation}
H=P^2/2m + B x^4 - A x^2 + \Lambda x \cos(\omega t),
\label{lbham}
\end{equation}
with $m=1$, $B=0.5$, $A=10$, $\Lambda=10$, $\omega=6.07$.  This
Hamiltonian has been used before in studies of quantum
chaos~\cite{linbal} and quantum decoherence~\cite{hsz} and, in the
parameter regime used, a substantial area of the accessible phase
space is stochastic.

\begin{figure}[here]
\begin{center}
\includegraphics[width=8.5cm,height=7.5cm]{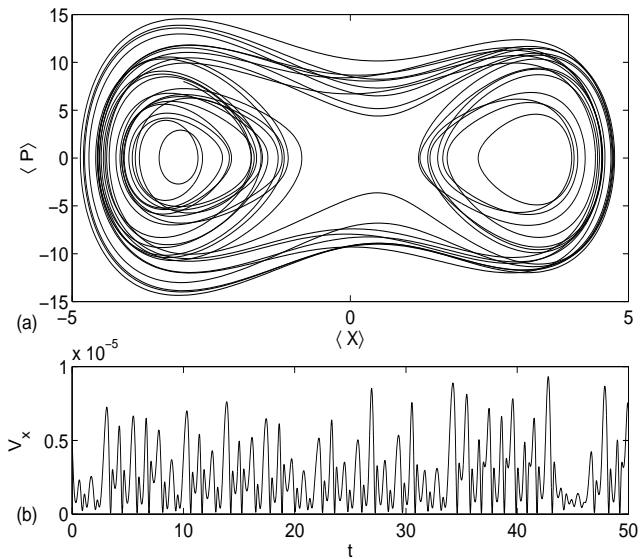}
\caption{(a) The quantum trajectory in phase space for a continuously
measured Duffing oscillator~\cite{bhj}, with $\hbar=10^{-5}$ and
$k=10^5$. (b) The position variance, $V_x$, as a function of
time. Note the smallness of the scale on the y-axis!}
\label{fig00}
\end{center}
\end{figure}

Numerical calculations at various values of $\hbar$ confirm that as
$\hbar$ is reduced, both the steady-state variance, and the resulting
noise (for optimal measurement strengths) are reduced, as expected.
As the dynamical time scale of this problem is $1 - 0.1$ (in units of
the driving period), the continuous observation record was averaged
over a period of $0.01$.  Similarly, as the range of the motion covers
distances of $O(10)$, we demand that the position be tracked to an
accuracy of $0.01$ to define an effective ``trajectory.''  To satisfy
this, we need to have $k \sim O(10^{5})$ or larger
[Cf. Eqn.~(\ref{kobs})]. In our example, we choose the energy to be
$O(10^2)$, the corresponding typical action turns out to be $O(10)$,
and the typical nonlinearity makes the RHS of Eqn.~\ref{klarge},
$O(1)$. We see that a choice of $\hbar = 10^{-5}$ and $k = 10^5$,
satisfies all the constraints for a classical motion.  In
Fig.~\ref{fig00} we demonstrate that in this regime, localization is
maintained along with low levels of trajectory
noise. Fig.~\ref{fig00}(a) shows a typical phase space trajectory,
with the position variance during the evolution, $V_x\equiv (\Delta
x)^2$, plotted in Fig.~\ref{fig00}(b). We find that the width $\Delta
x$ is always bounded by $3.4\times 10^{-3}$.  Furthermore, as is
immediately evident from the smoothness of the trajectory in
Fig.~\ref{fig00}(a), the noise is also negligible on these
scales. Additionally, one can verify that the quantum trajectory
evolution and that given by a classical trajectory with an equivalent
noise are essentially identical -- and chaotic -- yielding a Lyapunov
exponent of $0.57$. We return to discuss the Lyapunov exponent later
below.

\subsection{Chaos and the Weak Form of the QCT}

The weak form of the QCT utilizes the {\em coarse-grained}
distribution function (averaging over all measurements), whereas the
strong form refers to the {\em fine-grained} distribution for a single
measurement realization. It is important to reiterate that
nonexistence of the strong form of the QCT does not influence the
existence of the weak form of the QCT: It does not matter if the
distribution is too wide, as long as the classical and quantum
distributions agree, and, even if the backaction noise is large, the
coarse-grained distribution can remain smooth and the weak
quantum-classical correspondence still exist. Consequently, this
correspondence has to be approached in a different manner. In fact,
the weak version is just another way to state the conventional
decoherence idea~\cite{deco}; however, as discussed
elsewhere~\cite{shetal}, mere suppression of quantum interference does
not guarantee the QCT even in the weak form.

We now focus on a semiclassical analysis of the weak QCT for bounded,
classically chaotic open systems~\cite{ghss}. This analysis is best
regarded as a {\em regularization} of the singular $\hbar\rightarrow
0$ limit via the environmental interaction. This is distinct from the
state {\em localization} characteristic of the strong form of the
QCT. Given a small, but finite, value of $\hbar$, the aim is to
establish the existence of a timescale beyond which the dynamics of
open quantum and classical systems becomes statistically equivalent if
the environmental interaction is sufficiently strong. 

It has been demonstrated~\cite{ghss} that, for a bounded open system
with a classically chaotic Hamiltonian, the weak form of the QCT is
achieved by two parallel processes, both relying essentially on the
existence of environmental diffusion. First, the semiclassical
approximation for quantum dynamics, which breaks down for classically
chaotic systems due to overwhelming nonlocal interference, is
recovered as the environmental interaction filters these
effects. Second, environmental noise restricts the foliation of the
unstable manifold, the set of points which approach a hyperbolic point
in reverse time, allowing the semiclassical wavefunction to track this
modified classical geometry. In this way, the noise prevents classical
chaos from breaking the semiclassical approximation as
$\hbar\rightarrow 0$, and thus regularizes this limit. Note that this
approach explicitly incorporates both the stretching and folding
typical of hyperbolic regions as well as the role of the environment
as a filter on a phase-space quantum distribution.

We begin with a simple model of a quantum system weakly coupled to the
environment so as to maintain complete positivity for the subsystem
density matrix, $\rho(t)$, while subjecting it to a, time-local,
unitarity-breaking interaction.  These conditions mathematically
constrain the master equation to be of the so-called Lindblad form
~\cite{Lind}. If this environmental interaction couples to the
position, as is often the case, the master equation takes the
form: 
\begin{equation}
\frac{\partial f_{w}}{\partial t} =
L_{cl}f_{w}+L_{q}f_{w}+D\frac{\partial^{2}f_{w}}{\partial p^{2}},
\label{wme} 
\end{equation}
where $L_{cl}$, the classical Liouville operator, and $L_{q}$, the
quantum correction, can be easily identified from Eqn.~(\ref{qle}). We
note in passing that while the sum of $L_{cl}$ and $L_{q}$ is clearly
unitary, individually the operators are not unitary~\cite{shetal}. In
this simple master equation, we have neglected the dissipative
environmental channel and kept the diffusive channel for two reasons:
(i) the coupling to the environment is always assumed to be weak and
the dissipative timescales are, hence, very long, longer than the
dynamical timescales of interest, (ii) the weak form of the QCT arises
only from the diffusive channel, hence, dissipative effects are not of
interest here. 

When $L_{q}=0$, this equation reverts to the classical Fokker-Planck
equation. It is important to keep in mind that the specific form of
the diffusion coefficient depends strongly on the physical situation
envisaged. Thus, if the master equation describes a weakly coupled,
high temperature environment, $D=2m\gamma k_BT$ ($\gamma$ is the
damping coefficient)~\cite{Cald}, whereas for a weak, continuous
measurement of position, the diffusion due to quantum backaction is
$D=\hbar^2 k$~\cite{cmeqns}. The discussion below holds for all of
these cases.

Once the QCT occurs, the effects of $L_q$ in the evolution specified
by Eqn.~(\ref{wme}) are subdominant. Therefore, to understand how
environmental noise acts in this limit, it suffices to consider the
behavior of the corresponding classical Fokker-Planck equation. To do
this, it is convenient to examine the underlying Langevin equations
for noisy trajectories that unravel the evolution of the classical
distribution function when $L_{q}=0$. These are given by
\begin{eqnarray}
dq&=&\frac{p}{m} dt \nonumber\\
dp&=&f(q)dt+\sqrt{2D}dW
\label{langevin}
\end{eqnarray}
Using weak-noise perturbation theory, one can perform an expansion
about a hyperbolic fixed point and in this way obtain the spreading of
the position and momentum due to the diffusion. As a trajectory
evolves, it simultaneously smoothes over a transverse width in phase
space of size $\sqrt{{Dt}/(m{\lambda})}$ where $\lambda$ is the local
Lyapunov exponent~\cite{ghss}.

The smoothing implies a termination in the development of new phase
space structures at some finite time $t^{*}$, whose scaling behavior
can be determined. (Caveat: this need not be true in a non-compact
phase space.) The average motion of a trajectory is identical to its
deterministic motion, so that at time $t$, if the initial length in
phase space is $u_{0}$ ($u$ has units of square-root of phase-space
area), its current length will be approximately $u_{0}e^{\bar{\lambda}
t}$ as its forward time evolution will be dominated by its component
in the unstable direction.  Here $\bar{\lambda}$ is the time-averaged
positive Lyapunov exponent.  If the region is bounded within a phase
space area $A$, the typical distance between neighboring folds of the
trajectory is given by
\begin{equation}
l(t)\approx {A\over u_{0}}e^{-\bar{\lambda}t},
\label{dist}
\end{equation}
where $l(t)$ still carries the units of the square root of phase space
area.  However, since phase structures can only be known to within the
width specified above, the time at which any new structure will be
smoothed over is defined by  
\begin{equation} 
l(t^{*})\approx \sqrt{\frac{Dt^{*}}{m \bar{\lambda}}}.  
\label{nscale}
\end{equation}
The above two equations can be used to determine $t^*$, which only
weakly depends on $D$ and the prefactor in Eqn.~(\ref{dist}). Due to
the smoothing, one does not see an ergodic phase space region, but one
in which the large, short-time features that develop prior to $t^{*}$
are pronounced and the small, long-time features that develop later
are smoothed over by the averaging process.  Therefore, to approximate
noisy classical dynamics, a quantum system need not track all of the
fine scale structures, but only the larger features that develop
before the production of small scale structures terminates. 

To establish the conditions under which quantum dynamics can track
this modified phase space geometry, a semiclassical analysis can be
performed. In the Wigner function formalism, the breakdown of the
semiclassical approximation for chaotic systems can be associated with
an appealing geometric picture~\cite{BH,sh90} based on a uniform
approximation in phase space -- the Berry construction. We now use
this construction to understand how quantum interference in phase
space is smoothed over by the diffusion associated with environmental
coupling.

A general mixed state is an incoherent superposition of pure state
Wigner functions, where an individual semiclassical pure state Wigner
function can be formed by substituting the Van-Vleck semiclassical
wavefunction in Eqn.~(\ref{wdef}).  If we allow $q$ to be perturbed by
noise we can rewrite the classical action~\cite{Kos} 
\begin{equation}
S(q,t)\approx S(q_{C},t)-\sqrt{2D}\int_{0}^{t} dt \xi(t)q_{C}(t).
\end{equation}
Following Berry~\cite{BH}, we now rewrite the action for
the $i$th solution to the Hamilton-Jacobi equation as
\begin{eqnarray}
S_{i}(q_{C},t)&=&\int_{q_{C}(0)}^{q_{C}(t)}dq'p_{i}(q',t)-
\int_{0}^{t}dt'H(q_{C}(0),p_{i}(q_{C}(0),t')\nonumber\\
&\equiv&\int_{0}^{t}dt' {\cal H}_i(t'), 
\end{eqnarray}
where $p_{i}(q,t)$ is the $i$th branch of the momentum curve for a
given $q$. If we average over all noisy realizations, after separating
the contributions from identical branches, the following suggestive
expression for the noise averaged semiclassical Wigner function
obtains:
\begin{eqnarray} 
&&\frac{1}{2\pi\hbar}\int_{-\infty}^{\infty}dX
\exp\left(-{DtX^{2}\over 2\hbar^2}\right)
\bigg(\sum_{i}{\cal J}_{ii}\times\nonumber\\
&&\exp\bigg[\frac{i}{\hbar} 
\bigg\{\int_{\bar{q}_{-}}^{\bar{q}_{+}}dq'p_{i}(q',t)-pX\bigg\}\bigg]+
\nonumber\\
&&{2i}\sum_{i<j}{\cal J}_{ij}\sin\bigg[\frac{1}{\hbar}
\bigg\{\int_{q_{C}(0)}^{\bar{q}_{+}}dq'p_{i}(q',t)-
\int_{q_{C}(0)}^{\bar{q}_{-}}dq'p_{j}(q',t)\nonumber\\
&&-\int_{0}^{t}dt'\left({\cal H}_i-{\cal H}_j\right) +
\phi_{i}-\phi_{j}\bigg\}\bigg]\bigg);\\ 
\label{semiwig}
&&{\cal J}_{ij}\equiv
{C_{i}(\bar{q}_{+},t)C_{j}(\bar{q}_{-},t)\over 
\sqrt{|J_{i}(\bar{q}_{+},t)||J_{j}(\bar{q}_{-},t)|}}
\end{eqnarray} 
for Jacobian determinant $J_{i}(q,t)$ and transport coefficient
$C_{i}(q,t)$; $\bar{q}_{\pm}\equiv q\pm \frac{X}{2}$ and
$\phi_{i}=\pi\nu_{i}$, where $\nu_{i}$ is the $i$th Maslov
index~\cite{Maslov}. 

The dominant contributions to the integrals can be analyzed in
the stationary phase approximation~\cite{2}.  If $D=0$, these would
contribute phase coherences at values of $X$ that satisfy
$p_{i}(q+X/2,t)+p_{i}(q-X/2,t)-2pX=0$ for the first term in the
sum and $p_{i}(q+X/2,t)+p_{j}(q-X/2,t)-2pX=0$ for the second
term, the former being the famous Berry midpoint rule.  For a chaotic
system, Berry argued that, due to the proliferation of momentum
branches, $p_{i}(q,t)$, arising from the infinite number of foldings
of a bounded chaotic curve as $t\rightarrow\infty$, a semiclassical
approximation would eventually fail, since the interference fringes
stemming from a given $p_{i}$ could not be distinguished after a
certain time from those emanating from the many neighboring
branches~\cite{2}.  While the precise value of this time has since
been challenged numerically, the essential nature of this physical
argument has remained valid~\cite{HellTom}.

In the present case, however, the presence of noise acts as a
dynamical Gaussian filter, damping contributions for any solutions to
the above equation which are greater than $X\approx\hbar/\sqrt{Dt}$.
In other words, noise dynamically filters the long ``De Broglie''
wavelength contributions to the semiclassical integral, the very sort
of contributions which generally invalidate such an approximation.  If
we rescale the above result and combine it with our understanding of
how noise effects classical phase space structures, we can
qualitatively estimate whether or not a semiclassical picture is a
valid approximation to the dynamics.  As already discussed, $t^*$ is
the time when the formation of new classical structures ceases and
$l(t^*)$ is the associated scale over which classical structures are
averaged. The key requirement is then that the semiclassical phase
filters contributions of size
\begin{equation}
{\sqrt{\bar{\lambda} m}\hbar\over\sqrt{Dt^{*}}}\stackrel{<}{\sim}
l(t^{*}). 
\label{result}
\end{equation}
In other words, for a given branch, the phases with associated
wavelengths long enough to interfere with contributions from
neighboring branches are strongly damped, and the intuitive
semiclassical picture of classical phase-space distributions decorated
by local interference fringes recovered.

The weak form of the QCT is completed when the inequality
(\ref{result}) is satisfied. Substituting the scale of
classical smoothing (\ref{nscale}) in this inequality, we find
\begin{equation}
Dt^{*}\stackrel{>}{\sim} {\bar{\lambda} m \hbar}.
\label{tstar}
\end{equation}
[Note that the purely classical quantity $t^*$ is first independently
determined by solving Eqn.~(\ref{nscale}) and then compared to the
right side of the above equation.] While the left hand side of the
inequality contains the mutually dependent $t^*$ and $D$, the right
hand side depends only on fixed properties of the system and $\hbar$.
This condition, therefore, defines a threshold at which the
semiclassical approximation becomes stable and that may be set in
terms of either $D$ or $t^*$.  Once the threshold is met, $t^*$
becomes the time beyond which the semiclassical description is valid.
The semiclassical nature of this condition becomes more evident on
defining $S=l(t^{*})^2$ which, given that $l^2$ is an areal scale in
phase space for the diffusion averaged dynamics, has dimensions of
action.  A physical interpretation is more apparent on rewriting
(\ref{tstar}) as $S=l(t^*)^2 \stackrel{>}{\sim} \hbar$, which is
readily identified as the usual condition for the validity of a
semiclassical analysis.

The weak form of the QCT can also be demonstrated using the Duffing
example~\cite{ghss}. The dynamical evolution of the bounded motion is
dominated by the homoclinic tangle of a single hyperbolic fixed point.
As a result, the long-time chaotic evolution can be completely
characterized by the unstable manifold associated with that fixed
point~\cite{Guck}. The value of $\hbar$ is now set to $\hbar=0.1$,
significantly larger than when studying the strong form of the QCT.

\begin{figure}[here]
\begin{center}
\includegraphics[width=4in]{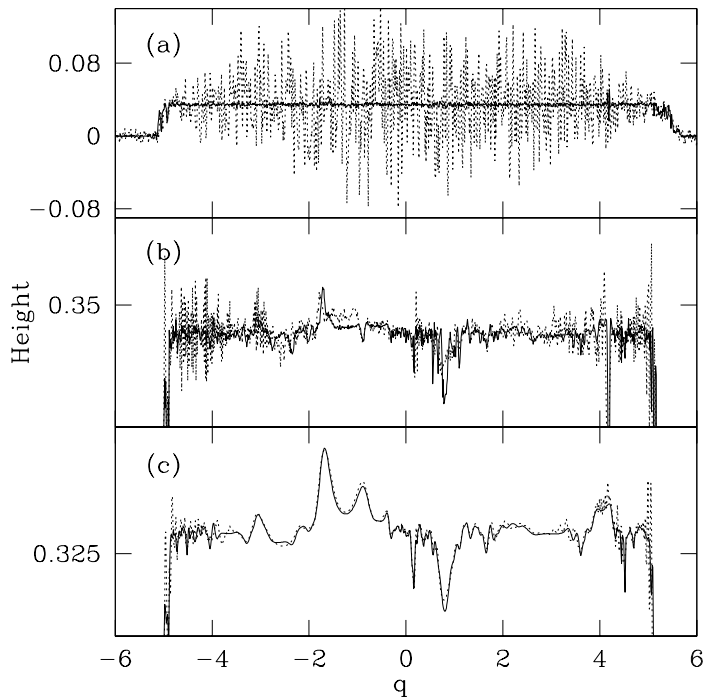}
\caption{\label{fig:SlicePlot} The weak form of the QCT for a driven
Duffing oscillator~\cite{ghss}: Sectional cuts of Wigner functions
(dashed lines) and classical distributions (solid lines), after 149
drive periods, taken at $p=0$ for diffusion coefficient values (a)
$D=10^{-5}$; (b) $D=10^{-3}$; (c) $D=10^{-2}$. Other parameter values
are stated in the text; the height is specified in scaled units.} 
\end{center} 
\end{figure}

The evolution of the corresponding distributions was numerically
calculated for both the classical and quantum master equations.
Fig.~\ref{fig:SlicePlot} shows sectional cuts at $p=0$ of the
quantum and classical phase space distribution functions for three
different values of the diffusion coefficient,
$D=10^{-5},~10^{-3},~10^{-2}$, after time $T=149$ evolution
periods. As already mentioned, $t^*$ varies slowly with $D$, and in
the three cases shown, $t^*$ ranges only from $\sim 20 - 14$ (note
that $t^*\ll T$). It is easy to check that the inequality
(\ref{tstar}) is strongly violated for $D=10^{-5}$, mildly violated
for $D=10^{-3}$, and approximately satisfied for $D=10^{-2}$. For
$D=10^{-5}$, the classical and quantum sections show no similarities,
as expected.  The quantum Wigner function also shows large negative
regions, reflecting strong quantum interference. On increasing $D$ to
$10^{-3}$ the magnitude of quantum coherence decreases dramatically
and the classical and quantum slices have the same average value, as
well as specific agreement on some large scale features.  The two
disagree, as expected, on the small scale structures.  This indicates
that, while the quantum and classical distributions do not exactly
match, the Wigner function has now become sensitive to the larger
features of the noise averaged classical distribution function,
indicative of the transition to a semiclassical regime.  At
$D=10^{-2}$, there is near perfect agreement between classical and
quantum distribution functions, save on the smallest scales.  When $D$
is of order unity, the inequalities enforcing the strong QCT at the
level of individual trajectories~\cite{bhj} are satisfied and the
agreement is essentially exact.  However, as indicated by
Fig.~\ref{fig:SlicePlot}(c), detailed agreement for quantum and
classical distribution functions can begin at much smaller values of
the diffusion constant.

\begin{figure}[here]
\begin{center}
\includegraphics[width=4in]{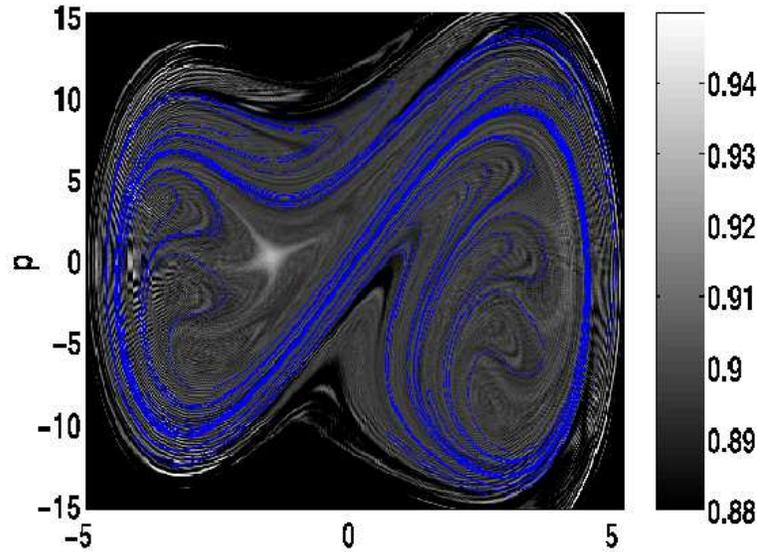}
\caption{\label{fig:WigMan}  Phase space rendering of the Wigner
function at time $t=149$ periods of driving~\cite{ghss}. The
early time part of the unstable manifold associated with the
noise-free dynamics is shown in blue. The value of $D=10^{-3}$ is not
sufficient to wipe out all the quantum interference which, as
expected, is most prominent near turns in the manifold.}  
\end{center}
\end{figure}

For more detailed evidence that, at $D=10^{-3}$, one is entering a
semiclassical regime, in Fig.~\ref{fig:WigMan} we superimpose an image
of the large scale features of the classical unstable manifold on top
of the full quantum Wigner distribution at $D=10^{-3}$ after 149 drive
periods [case (b) of Fig.~2].  The quantum phase space clearly
exhibits local interference fringes around the large lobe-like
structures associated with the short-time evolution of the unstable
manifold.  The appearance of local fringing about classical structures
is direct evidence of a semiclassical evolution, where interference
effects appear locally around the backbone of a classical evolution.
This is in sharp contrast to the global diffraction pattern seen for
$D=0$, where the contributions from individual curves cannot be
distinguished, suppressing the appearance of any classical
structure~\cite{hsz}.

\section{Chaos in Quantum Mechanics}

At this point, our analysis of measured quantum dynamical systems may
be said to have harmonized quantum and classical mechanics in the
sense that the strong and weak forms of the QCT have appeared
naturally. While this is certainly pleasing, we wish to go further and
ask whether the formalism can be tested by making predictions that are
experimentally verifiable and depend uniquely on the nonlinear nature
of the conditioned evolution. One very interesting idea is the
real-time control of quantum systems using state-estimation as
pioneered by Belavkin~\cite{dhjmt} or direct feedback of the measured
classical current~\cite{wm}. Although quantum feedback control
applications~\cite{qfcapp} have their own importance, we now return to
the original burning question: Is there chaos in quantum mechanics?

In a limiting case, the answer is clearly in the affirmative. We have
already shown that quantum distributions, provided certain conditions
are met, can evolve while staying localized and be only very weakly
perturbed by noise. In the classical limiting case, we recover
localized classical trajectories, and these can certainly be
chaotic. But what if these conditions are not satisfied?

This is the question addressed and answered in Ref.~\cite{hjs}. By
defining and computing the Lyapunov exponent for an observed quantum
system deep in the quantum regime, we were able to show that the
system dynamics is chaotic. Further, the Lyapunov exponent is not the
same as that of the classical dynamics that emerges in the classical
limit. Since the quantum system in the absence of measurement is not
chaotic, this chaos must emerge as the strength of the measurement is
increased, and we examined the nature of this emergence.

To do this, we must first make certain that we can quantify
the existence of chaos in a robust way. The rigorous quantifier of
chaos in a dynamical system is the maximal Lyapunov
exponent~\cite{ER}. The exponent yields the (asymptotic) rate of
exponential divergence of two trajectories which start from
neighboring points in phase space, in the limit in which they evolve
to infinity, and the neighboring points stay infinitesimally close. The
maximal Lyapunov exponent characterizes the sensitivity of the system
evolution to changes in the initial condition: if the exponent is
positive, then the system is exponentially sensitive to initial
conditions, and is said to be chaotic. We now discuss how this notion
can be applied to observation-conditioned evolution of quantum
expectation values.

A single quantum mechanical particle is in principle an infinite
dimensional system. However, for the purpose of defining an
observationally relevant Lyapunov exponent, it is sufficient to use a
single projected data stream: let us consider the expectation value of
the position, $\langle x(t)\rangle$. The important quantity is thus
the divergence, $\Delta(t) = |\langle x(t)\rangle - \langle
x_{\mbox{\scriptsize fid}}(t) \rangle |$, between a fiducial
trajectory and a second trajectory infinitesimally close to it. It is
important to keep in mind that the system is driven by noise. Since we
wish to examine the sensitivity of the system to changes in the
initial conditions, and not to changes in the noise, we must hold the
noise realization fixed when calculating the divergence. The Lyapunov
exponent is thus
\begin{equation}
  \lambda \equiv \lim_{t\rightarrow\infty} 
  \lim_{\Delta_s(0)\rightarrow 0} \frac{\ln \Delta_s (t)}{t}
  \equiv \lim_{t\rightarrow\infty}\lambda_s(t)
\label{Lyap} 
\end{equation} 
where the subscript $s$ denotes the noise realization. This definition
is the obvious generalization of the conventional ODE definition to
dynamical averages, where the noise is treated as a drive on the
system. Indeed, under the conditions when (noisy) classical motion
emerges, and thus when localization holds, it reduces to the
conventional definition, and yields the correct classical Lyapunov
exponent. To combat slow convergence, we measure the Lyapunov exponent
by averaging over an ensemble of finite-time exponents $\lambda_s(t)$
instead of taking the asymptotic long-time limit for a single
trajectory.

A key result now follows: In unobserved, i.e., isolated quantum
dynamical systems, it is possible to prove, by employing unitarity and
the Schwarz inequality, that $\lambda$ vanishes; the finite-time
exponent, $\lambda(t)$, decays away as $1/t$~\cite{tocome}. From the
Kosloff-Rice theorem we know, of course, that the Lyapunov exponent
must be zero, since the overall evolution is integrable, but this
result gives us a quantitative statement regarding the decay of the
exponent. It turns out that this particular result applies also to the
evolution of averages in isolated classical systems and, in this
sense, is more general than Kosloff-Rice. As we have emphasized
earlier, once measurement is included, the evolution becomes nonlinear
and the Lyapunov exponent need not vanish classically or quantum
mechanically.

\begin{figure}[here]
\begin{center}
   \includegraphics[width=8.5cm,height=6.5cm]{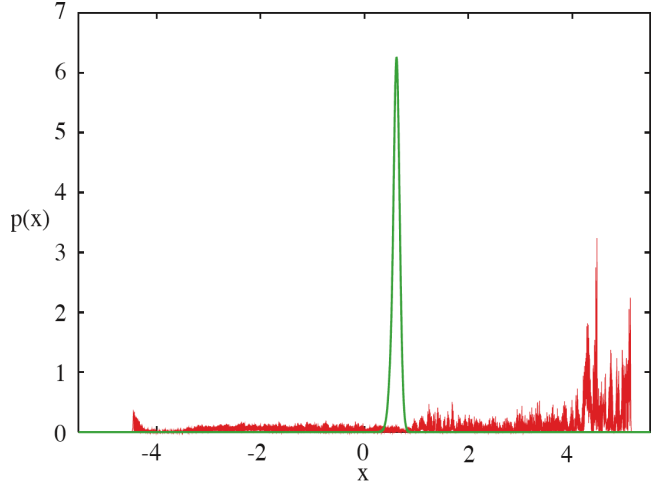}
   \caption[example]{ \label{fig0} Position distribution for
     the Duffing oscillator with measurement strengths $k=0.01$ (red)
     and $k=10$ (green), demonstrating measurement-induced
     localization ($k=10$) as the measurement coupling is
     increased~\cite{hjs}. The momentum distribution behaves
similarly.} 
\end{center}
\end{figure}

As a particular system of interest, we turn once again to the Duffing
oscillator, this time with $\hbar=10^{-2}$, which is small enough so
that the system makes a transition to classical dynamics when the
measurement is sufficiently strong. As we increase the measurement
strength, we can examine the transformation from essentially isolated
quantum evolution all the way to the known chaos of the classical
Duffing oscillator. To examine the emergence of chaos, in
Ref.~\cite{hjs} we solved for the evolution of the system for $k=
5\times 10^{-4}, 10^{-3}, 0.01, 0.1, 1, 10$. When $k\leq 0.01$, the
distribution is spread over the entire accessible region, and
Ehrenfest's theorem is not satisfied. Conversely, for $k=10$, the
distribution is well-localized (Fig.~\ref{fig0}), and Ehrenfest's
theorem holds throughout the evolution. Since the backaction noise,
characterized by the momentum diffusion coefficient, $D=\hbar^2k$,
remains small, at this value of $k$ the motion is that of the
classical system, to a very good approximation.

\begin{figure}[here]
\begin{center}
   \includegraphics[width=9.5cm,height=8.5cm]{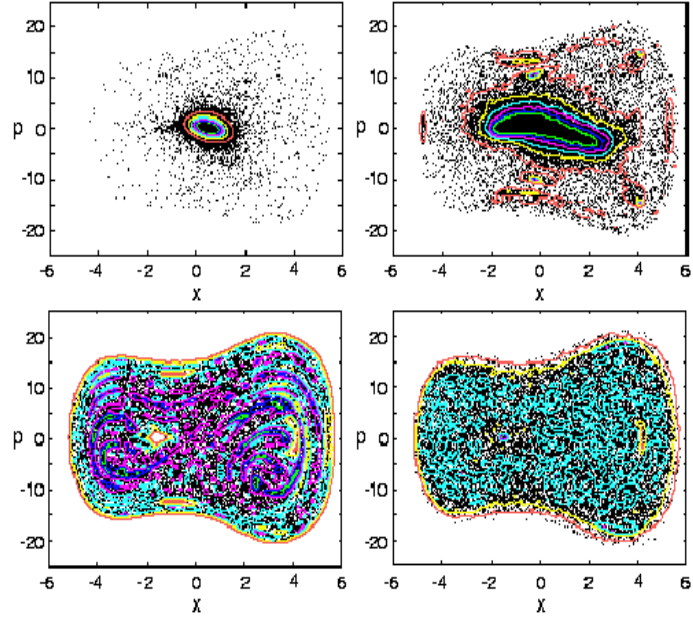}
   \caption[example]{ \label{fig1} Phase space stroboscopic
maps~\cite{hjs} for the observed Duffing oscillator for 4 different
measurement strengths, $k=5\times 10^{-4}$, 0.01 (top), and 1, 10
(bottom). Contour lines are superimposed to provide a measure of local
point density at relative density levels of
$0.05,~0.15,~0.25,~0.35,~0.45,$ and $0.55$.} 
\end{center}  
\end{figure} 

Stroboscopic maps help reveal the global structural transformation in
phase space in going from quantum to classical dynamics
(Fig.~\ref{fig1}). The maps consist of points through which the system
passes at time intervals separated by the period of the driving
force. For very small $k$, $\langle x\rangle$ and $\langle p\rangle$
are largely confined to a region in the center of phase
space. Somewhat remarkably, at $k=0.01$, although the system is
largely delocalized, as shown in Fig.~\ref{fig0}, nontrivial structure
appears, with considerable time being spent in certain outer
regions. By $k=1$ the localized regions have formed into narrower and
sharper swirling coherent structures. At $k=10$ the swirls disappear,
and we retrieve the uniform chaotic sea of the classical map (the
small ``holes'' are periodic islands). The swirls in fact correspond to
the unstable manifolds of the classical motion. Classically, these
manifolds are only visible at short times, as continual and repeated
folding eventually washes out any structure in the midst of a uniform
tangle.  In the quantum regime, however, the weakness of the
measurement, with its inability to crystallize the fine structure, has
allowed them to survive: we emphasize that the maps result from
long-time integration, and are therefore essentially time-invariant.

To calculate the Lyapunov exponent we implemented a numerical version
of the classical linearization technique~\cite{wolf}, suitably
generalized to quantum trajectories. The method was tested on a
classical noisy system with comparison against results obtained from
solving the exact equations for the Lyapunov
exponents~\cite{habibryne}. The calculation is very numerically
intensive, as it involves integrating the stochastic Schrodinger
equation equivalent to the SME (\ref{condq}) over thousands of driving
periods, and averaging over many noise realizations; parallel
supercomputers were invaluable for this task.

\begin{figure}[here]
\begin{center}
   \includegraphics[width=8.5cm,height=10cm]{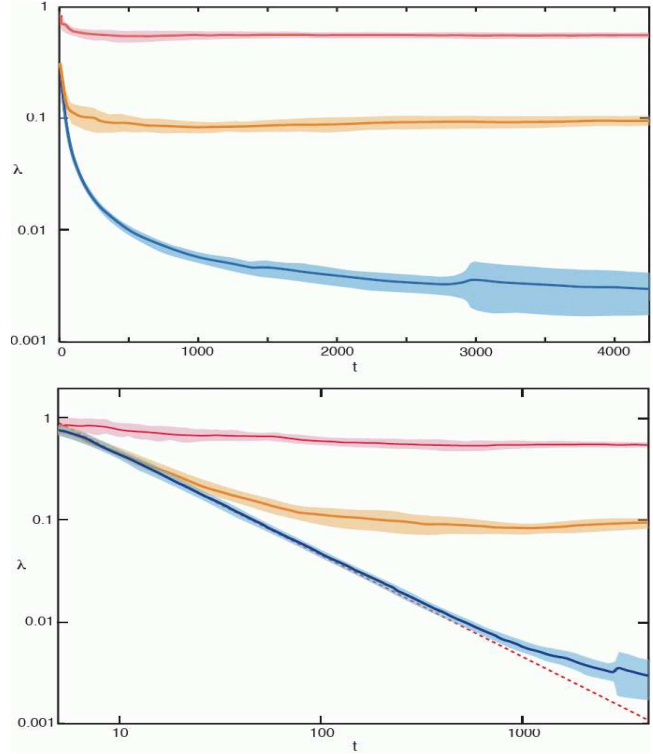}
   \caption[example]{ \label{fig2} Finite-time Lyapunov exponents
$\lambda(t)$ for measurement strengths $k=5\times 10^{-4},~0.01,~10$,
averaged over 32 trajectories for each value of $k$ (linear
scale in time, top, and logarithmic scale, bottom; bands indicate
the standard deviation over the 32 trajectories)~\cite{hjs}. The
(analytic) $1/t$ fall-off at small $k$ values, prior to the asymptotic
regime, is evident in the bottom panel. The unit of time is the
driving period.}    
\end{center}
\end{figure}

\begin{figure}[here]
\begin{center}
   \includegraphics[width=8.5cm,height=6.5cm]{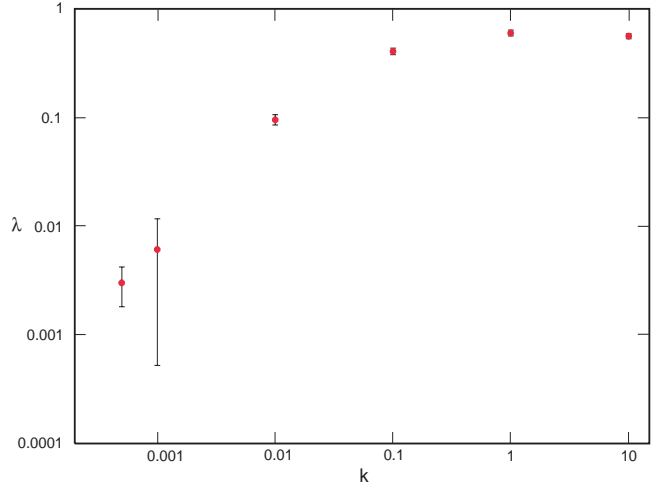}
   \caption[example]{ \label{fig3} The emergence of chaos~\cite{hjs}:
The Lyapunov exponent $\lambda$ as a function of measurement strength
$k$. Error-bars follow those of Fig.~6, taken at the final time.} 
\end{center}
\end{figure}

The computations show that as $t$ is increased, for nonzero $k$, the
value obtained for $\lambda(t)$ falls as $1/t$, following the behavior
expected for $k=0$, until a point at which an asymptotic regime takes
over, stabilizing at a finite value of the Lyapunov exponent as
$t\rightarrow\infty$. This behavior is shown in Fig.~\ref{fig2} for
three values of $k$. The Lyapunov exponent as a function of $k$ is
shown in Fig.~\ref{fig3}. The exponent increases over two orders of
magnitude in an approximately power-law fashion as $k$ is varied from
$5\times 10^{-4}$ to $10$, before settling to the classical value,
$\lambda_{Cl}=0.57$. The results in Figs.~\ref{fig2} and \ref{fig3}
show clearly that chaos emerges in the observed quantum dynamics well
before the limit of classical motion is obtained.

We also computed the Lyapunov exponent for the quantum system when its
action is sufficiently small that smooth classical dynamics cannot
emerge, even for strong measurement~\cite{hjs}. Taking a value of
$\hbar=16$, we find that for $k=5\times 10^{-3}$, $\lambda=0.029\pm
0.008$, for $k=0.01$, $\lambda=0.046\pm 0.01$ and for $k=0.02$,
$\lambda=0.077\pm 0.01$. Thus the system is once again chaotic, and
becomes more strongly chaotic the more strongly it is observed. From
these striking results, it is clear that there exists a purely {\em
quantum} regime in which an observed system, while behaving in a
fashion quite distinct from its classical limit, nevertheless evolves
chaotically with a finite Lyapunov exponent, also distinct from the
classical value.

It is worth pointing out that an analogous analysis can also be
carried out for a continuously observed classical system. As mentioned
previously, an {\em unobserved} probabilistic classical system also
has provably zero Lyapunov exponent: the average of $x$ for an
ensemble of classical particles does not exhibit chaos, due to the
linearity of the Liouville equation~\cite{tocome}. If we consider a
noiseless observed chaotic classical system -- possible since
classical measurements are by definition passive (no backaction noise)
-- then even the weakest meaningful measurement will, over time,
localize the probability density, generating an effective trajectory
limit, and thus the classical Lyapunov exponent,
$\lambda_{Cl}$~\cite{tocome}. Noise can always be be injected into
classical systems as an external drive, nevertheless, in the limit of
weak noise, the system will once again possess the noiseless exponent
$\lambda_{Cl}$: In a classical system the external noise is not
connected to the strength of the measurement, so one can
simultaneously have strong measurement and weak noise, which, as we
have seen, is possible in the quantum theory only under specific
conditions~\cite{bhj}.

As one way to understand the classical case, we can employ the quantum
result as an intermediate step. Consider the quantum Lyapunov exponent
at a fixed value of $k$ (where $\lambda < \lambda_{Cl}$) as in
Fig.~\ref{fig3}. If the value of $\hbar$ is now reduced, the dynamics
of the system must tend to the classical limit as the
quantum-classical correspondence inequalities of Ref.~\cite{bhj} are
better satisfied. Thus the Lyapunov exponent in the classical limit of
quantum theory -- which, to a very good approximation, is just
classical dynamics driven by weak noise -- must tend to
$\lambda_{Cl}$. If, however, the noise is not weak, an observed
classical system, like a quantum system outside the classical regime,
will also not be localized, and may well have an exponent different
from $\lambda_{Cl}$. In addition, one may expect the non-localized
quantum and classical evolutions to have quite different Lyapunov
exponents, especially when $\hbar$ is large on the scale of the phase
space, as quantum and classical evolutions generated by a given
nonlinear Hamiltonian are essentially different~\cite{shetal}.  The
nature of the Lyapunov exponent for non-localized classical systems,
and its relationship to the exponent for quantum systems is a very
interesting open question.

\section{Concluding Remarks}

To summarize, we have presented a simple analysis of continuously
observed classical and quantum dynamical systems. This analysis is in
fact required to deal with next-generation experiments and underlies
the nascent field of real-time quantum feedback control. Major results
include an intuitive and quantitative understanding of the
quantum-classical transition. It is pleasing that both the strong and
weak forms of the QCT can eventually be understood as a macroscopic
limit of observed-system quantum mechanics, i.e., whenever the
observed system action $S\gg\hbar$.

Perhaps, most interestingly, we have obtained clear predictions for
dynamical chaos in observed quantum systems that are far from the
classical regime. We emphasize that the chaos identified here is not
merely a formal result -- even deep in the quantum regime, the
Lyapunov exponent can be obtained from measurements on a real system
as in near-future cavity QED and nanomechanics
experiments~\cite{exp}. Experimentally, one would use the known
measurement record to integrate the SME (\ref{condq}); this provides
the time evolution of the mean value of the position. From this
fiducial trajectory, given the knowledge of the system Hamiltonian,
the Lyapunov exponent can be obtained by following the procedure
described here.

\begin{acknowledgments}

SH thanks the organizers of the 16th Florida Workshop in Nonlinear
Astronomy and Physics, dedicated to the memory of Henry Kandrup, for
their kind invitation to lecture at the meeting. Large-scale parallel
computing support from Los Alamos National Laboratory's Institutional
Computing Initiative is gratefully acknowledged. This research is
supported by the Department of Energy, under contract W-7405-ENG-36.

\end{acknowledgments}

\end{document}